\documentclass[twocolumn]{aastex63}

\usepackage{upgreek}
\usepackage{mathtools}
\usepackage{graphicx}
\usepackage{color}
\usepackage{float}
\usepackage{xspace}
\usepackage{booktabs} 
\usepackage{txfonts}
\usepackage{soul}

\usepackage{natbib}

\newcommand{\frb}{FRB~20201124A\xspace}

\shorttitle{}
\shortauthors{Nimmo et al. }

\makeatother

\begin{document}

\title{Milliarcsecond localisation of the repeating \frb}

\author[0000-0003-0510-0740]{K.~Nimmo}
  \affiliation{ASTRON, Netherlands Institute for Radio Astronomy, Oude Hoogeveensedijk 4, 7991 PD Dwingeloo, The Netherlands}
  \affiliation{Anton Pannekoek Institute for Astronomy, University of Amsterdam, Science Park 904, 1098 XH, Amsterdam, The Netherlands}
\author[0000-0002-5794-2360]{D.~M.~Hewitt}
  \affiliation{Anton Pannekoek Institute for Astronomy, University of Amsterdam, Science Park 904, 1098 XH, Amsterdam, The Netherlands}
\author[0000-0003-2317-1446]{J.~W.~T.~Hessels}
  \affiliation{Anton Pannekoek Institute for Astronomy, University of Amsterdam, Science Park 904, 1098 XH, Amsterdam, The Netherlands}
  \affiliation{ASTRON, Netherlands Institute for Radio Astronomy, Oude Hoogeveensedijk 4, 7991 PD Dwingeloo, The Netherlands}
\author[0000-0001-6664-8668]{F.~Kirsten}
  \affiliation{ASTRON, Netherlands Institute for Radio Astronomy, Oude Hoogeveensedijk 4, 7991 PD Dwingeloo, The Netherlands}
  \affiliation{Department of Space, Earth and Environment, Chalmers University of Technology, Onsala Space Observatory, SE-439 92 Onsala, Sweden}
\author[0000-0001-9814-2354]{B.~Marcote}
  \affiliation{Joint Institute for VLBI ERIC (JIVE), Oude Hoogeveensedijk 4, 7991PD Dwingeloo, The Netherlands}
\author[0000-0002-7722-8412]{U.~Bach}
  \affiliation{Max-Planck-Institut f\"ur Radioastronomie, Auf dem H\"ugel 69, 53121 Bonn, Germany}
\author[0000-0003-1771-1012]{R.~Blaauw}
  \affiliation{ASTRON, Netherlands Institute for Radio Astronomy, Oude Hoogeveensedijk 4, 7991 PD Dwingeloo, The Netherlands}
\author[0000-0002-8265-4344]{M.~Burgay}
  \affiliation{Istituto Nazionale di Astrofisica, Osservatorio Astronomico di Cagliari, Via della Scienza 5, I-09047, Selargius, Italy}
\author[0000-0002-5924-3141]{A.~Corongiu}
  \affiliation{Istituto Nazionale di Astrofisica, Osservatorio Astronomico di Cagliari, Via della Scienza 5, I-09047, Selargius, Italy}
\author[0000-0002-9812-2078]{R.~Feiler}
  \affiliation{Institute of Astronomy, Faculty of Physics, Astronomy and Informatics, Nicolaus Copernicus University, Grudziadzka 5, 87-100 Toru\'n, Poland}
\author[0000-0003-4056-4903]{M.~P.~Gawro\'nski}
  \affiliation{Institute of Astronomy, Faculty of Physics, Astronomy and Informatics, Nicolaus Copernicus University, Grudziadzka 5, 87-100 Toru\'n, Poland}
\author[0000-0002-8657-8852 ]{M.~Giroletti}
  \affiliation{Istituto Nazionale di Astrofisica, Istituto di Radioastronomia, Via Gobetti 101, I-40129, Bologna, Italy}
\author[0000-0002-5307-2919]{R.~Karuppusamy}
  \affiliation{Max-Planck-Institut f\"ur Radioastronomie, Auf dem H\"ugel 69, 53121 Bonn, Germany}
\author[0000-0002-5575-2774]{A.~Keimpema}
  \affiliation{Joint Institute for VLBI ERIC (JIVE), Oude Hoogeveensedijk 4, 7991PD Dwingeloo, The Netherlands}
\author[0000-0002-0321-8588]{M.~A.~Kharinov}
  \affiliation{Institute of Applied Astronomy, Russian Academy of Sciences, Kutuzova Embankment 10, St. Petersburg, 191187, Russia}
\author[0000-0002-3669-0715 ]{M.~Lindqvist}
  \affiliation{Department of Space, Earth and Environment, Chalmers University of Technology, Onsala Space Observatory, SE-439 92 Onsala, Sweden}
\author[0000-0002-1482-708X]{G.~Maccaferri}
  \affiliation{Istituto Nazionale di Astrofisica, Istituto di Radioastronomia, Via Gobetti 101, I-40129, Bologna, Italy}
\author[0000-0002-8466-7026]{A.~Melnikov}
  \affiliation{Institute of Applied Astronomy, Russian Academy of Sciences, Kutuzova Embankment 10, St. Petersburg, 191187, Russia}
\author[0000-0002-3355-2261]{A.~Mikhailov}
  \affiliation{Institute of Applied Astronomy, Russian Academy of Sciences, Kutuzova Embankment 10, St. Petersburg, 191187, Russia}
\author[0000-0001-9381-8466]{O.~S.~Ould-Boukattine}
  \affiliation{Anton Pannekoek Institute for Astronomy, University of Amsterdam, Science Park 904, 1098 XH, Amsterdam, The Netherlands}
\author[0000-0002-5195-335X]{Z.~Paragi}
  \affiliation{Joint Institute for VLBI ERIC (JIVE), Oude Hoogeveensedijk 4, 7991PD Dwingeloo, The Netherlands}
\author[0000-0001-7397-8091]{M.~Pilia}
  \affiliation{Istituto Nazionale di Astrofisica, Osservatorio Astronomico di Cagliari, Via della Scienza 5, I-09047, Selargius, Italy}
\author[0000-0001-5902-3731]{A.~Possenti}
  \affiliation{Istituto Nazionale di Astrofisica, Osservatorio Astronomico di Cagliari, Via della Scienza 5, I-09047, Selargius, Italy}
\author[0000-0001-6170-2282]{M.~P.~Snelders}
  \affiliation{Anton Pannekoek Institute for Astronomy, University of Amsterdam, Science Park 904, 1098 XH, Amsterdam, The Netherlands}
\author[0000-0003-2775-442X]{G.~Surcis}
  \affiliation{Istituto Nazionale di Astrofisica, Osservatorio Astronomico di Cagliari, Via della Scienza 5, I-09047, Selargius, Italy}
\author{M.~Trudu}
  \affiliation{Istituto Nazionale di Astrofisica, Osservatorio Astronomico di Cagliari, Via della Scienza 5, I-09047, Selargius, Italy}
\author[0000-0002-8476-6307]{T.~Venturi}
  \affiliation{Istituto Nazionale di Astrofisica, Istituto di Radioastronomia, Via Gobetti 101, I-40129, Bologna, Italy}
\author[0000-0002-2700-9916]{W.~Vlemmings}
  \affiliation{Department of Space, Earth and Environment, Chalmers University of Technology, Onsala Space Observatory, SE-439 92 Onsala, Sweden}
\author[0000-0002-9786-8548]{N.~Wang}
  \affiliation{Xinjiang Astronomical Observatory, 150 Science 1-Street, Urumqi, Xinjiang 830011, China }
\author[0000-0002-2322-5232]{J.~Yang}
  \affiliation{Department of Space, Earth and Environment, Chalmers University of Technology, Onsala Space Observatory, SE-439 92 Onsala, Sweden}
\author{J.~Yuan}
  \affiliation{Xinjiang Astronomical Observatory, 150 Science 1-Street, Urumqi, Xinjiang 830011, China }
\correspondingauthor{K.~Nimmo}\email{k.nimmo@uva.nl}

\begin{abstract}
Very long baseline interferometric (VLBI) localisations of repeating fast radio bursts (FRBs) have demonstrated a diversity of local environments: from nearby star-forming regions to globular clusters. Here we report the VLBI localisation of \frb using an {\it ad-hoc} array of dishes that also participate in the European VLBI Network (EVN). In our campaign, we detected 18 total bursts from \frb\ at two separate epochs. By combining the visibilities from both observing epochs, we were able to localise \frb\ with a 1-$\sigma$ error of 4.5\,milliarcseconds (mas). We use the relatively large burst sample to investigate astrometric accuracy, and find that for $\gtrsim20$ baselines ($\gtrsim7$ dishes) that we can robustly reach milliarcsecond precision even using single-burst data sets.  Sub-arcsecond precision is still possible for single bursts, even when only $\sim$ six baselines (four dishes) are available.  We explore two methods for determining the individual burst positions: the peaks of the dirty maps and a Gaussian fit to the cross fringe pattern on the dirty maps. We found the latter to be more reliable due to the lower mean and standard deviation in the offsets from the FRB position. Our VLBI work places \frb\ 705$\pm$26\,mas (1-$\sigma$ errors) from the optical centre of the host galaxy, and consistent with originating from within the recently-discovered extended radio structure associated with star-formation in the host galaxy. Future high-resolution optical observations, e.g. with {\em Hubble Space Telescope}, can determine the proximity of our \frb\ VLBI position to nearby knots of star formation.
\end{abstract}

%

\section{Introduction}\label{sec:intro}

Fast radio bursts (FRBs) are highly luminous, short-duration coherent radio transients \citep[for recent reviews see][]{petroff_2019_aarv,petroff_2021_arxiv,cordes_2019_araa}. The vast majority of the observed FRB population are apparently one-off events, but a few percent of the known FRBs have been seen to repeat. Whether all FRBs are capable of repeating, or if the observed FRB population comes from multiple origins, remains debated.  Nonetheless, the large sample provided by CHIME/FRB \citep{chime_2018_apj} shows statistical differences between the properties of apparent one-offs and repeaters \citep{pleunis_2021_arxiv}.  The repeating sources are particularly valuable in our efforts to understand the nature of FRBs, since they allow for follow-up observations to, e.g., explore their burst energy distribution \citep{gourdji_2019_apjl,li_2021_natur}, characterise the evolution of the burst properties with time and frequency \citep[][]{gajjar_2018_apj,michilli_2018_natur,hilmarsson_2021_apjl,pleunis_2021_apjl}, and probe the immediate surroundings of the FRB source through precise localisation with very long baseline interferometry (VLBI) and high-resolution optical imaging \citep{bassa_2017_apjl,tendulkar_2021_apjl,mannings_2021_apj}.

To date, 19 FRBs have been localised with sufficient precision ($<$\, a few arcseconds) to identify their host galaxy \citep[e.g.,][]{chatterjee_2017_natur,ravi_2019_natur,bannister_2019_sci,bhandari_2021_arxiv}\footnote{\url{https://frbhosts.org/}}.  Thus far, only three of these are localised to 1--10 milliarcsecond precision using VLBI \citep{marcote_2017_apjl,marcote_2020_natur,kirsten_2021_arxiv}.  With VLBI precision, the exact location of the FRB progenitor within the host galaxy can be determined and matched with any prominent features, like spiral arms, the galactic nucleus, or star-forming regions.  This is particularly useful if the host can be resolved with \emph{Hubble Space Telescope} observations \citep{bassa_2017_apjl,tendulkar_2021_apjl,mannings_2021_apj}.  

VLBI localisations of repeating FRBs have uncovered a diversity of local environments: FRB~20121102A, the first-known repeater, is spatially coincident with a compact persistent radio source and inside a star-forming region in a dwarf host galaxy \citep{chatterjee_2017_natur,marcote_2017_apjl,tendulkar_2017_apjl,bassa_2017_apjl}; FRB~20180916B is located in the spiral arm of a Milky Way-like host, approximately 250\,pc from a prominent star-forming knot \citep{marcote_2020_natur,tendulkar_2021_apjl}; and FRB~20200120E was discovered to be in a globular cluster that is associated with the grand design spiral galaxy M81 \citep{bhardwaj_2021_apjl,kirsten_2021_arxiv}. This diversity highlights that the progenitor of repeating FRBs is able to live in different types of environments or, alternatively, that there is more than one progenitor type for repeating FRBs. Additional VLBI localisations are key to finding similarities between FRBs and other astronomical source classes, and to further explore the full diversity of FRB local environments.  

In this Letter, we present the European VLBI Network (EVN) localisation of the repeating \frb. \frb\ entered a period of heightened activity in April 2021 \citep{chime_2021_atel}, during which we detected a total of 18 bursts at two epochs. A preliminary EVN position was originally reported in \citet{marcote_2021_atel}, who used data only from the first epoch. Here, we describe in detail the interferometric observations (\S\,2), followed by the analysis and results (\S\,3).  Because we detected a relatively large sample of bursts at two separate epochs, a first for FRB observations using the VLBI technique, we use this to test the astrometric accuracy of the method (\S\,4).  We investigate the astrometric accuracy for individual bursts and low number of available antennas.  Finally, we discuss our results in the context of recent studies of \frb\ and other FRBs.  

\section{Observations}

We observed \frb\ on April 10 2021 15:00 -- 21:00 UT (project ID pr153a) and April 19 2021 13:30 -- 19:30 UT (project ID pr156a) as part of our ongoing FRB VLBI localisation project, PRECISE (Pinpointing REpeating ChIme Sources with Evn dishes)\footnote{\url{https://www.ira.inaf.it/precise/Home.html}}. We used an {\it ad-hoc} array of $6$ and $9$ radio telescopes (that are also part of the EVN) on April 10 and 19, respectively.  We pointed at the position derived from earlier Very Large Array (VLA) observations: RA (J2000) $= 5^{\rm h}08^{\rm m}03.5^{\rm s}$, Dec (J2000) $= +26^\circ03^\prime37.8^{\prime\prime}$ \citep{law_2021_atel}. The telescopes that participated in our campaign were Onsala (On-85), Toru\'n, Irbene, Westerbork single-dish RT1, Noto and Effelsberg at Epoch\,1, and Onsala (On-85), Toru\'n, Irbene, Westerbork single-dish RT1, Medicina, Svetloe, Badary, Sardinia and Effelsberg at Epoch\,2. The central observing frequency of our observations was 1.4\,GHz, and the bandwidth and number of subbands was dependent on the capabilities at each station (see Table\,\ref{tab:tele}). Phase referencing was done with a cycle time of $6$--$8.5$\,min: $1.5$--$2.5$\,min on the phase calibrator source (J0502$+$2516, at a separation of $1.4^\circ$ from \frb), followed by $4.5$--$6.5$\,min on target. We use the position of J0502$+$2516 reported in the rfc2018a catalogue\footnote{\url{http://astrogeo.org/vlbi/solutions/rfc\_2018a/}} during correlation: RA (J2000) $= 5^{\rm h}02^{\rm m}58.474768^{\rm s}$, Dec (J2000) $= +25^\circ16^\prime25.27549^{\prime\prime}$ (offset by 0.09\,mas in RA and Dec from the rfc2021c catalogue\footnote{\url{http://astrogeo.org/sol/rfc/rfc_2021c/}}, i.e. within the listed uncertainty of $0.12$~mas). Scans of J1048$+$7143 and J0555$+$3948 were taken to use as fringe finders and bandpass calibrators in the first and second epoch, respectively. Additionally, we observed the pulsar PSR~J2257$+$5909 for 5\,min per epoch to test the data quality of the single-dish data, and frequent, phase-referenced scans of J0501$+$2530 to use as an interferometric check source. In total, we observed the target \frb\ for 7.1\,hours.

We recorded raw voltage data with dual circular polarisation and 2-bit sampling from each telescope that participated, in either VDIF \citep{whitney_2010_ivs} or MARK5B (in the case of Svetloe and Badary; \citealt{whitney_2004_evn}) format. In parallel, we simultaneously recorded pulsar backend data at Effelsberg during both epochs and at Sardinia during Epoch 2. At Effelsberg, total intensity filterbank data were recorded using the PSRIX pulsar backend \citep{lazarus_2016_mnras}. Unfortunately, the PSRIX data recorded on April 10 suffered from an incorrect observing set-up and were not usable. The PSRIX data on April 19 was recorded with time and frequency resolutions of 102.4\,$\upmu$s and 0.49\,MHz, respectively, and a usable frequency range of $1255$--$1505$\,MHz.  At Sardinia, the pulsar data were recorded using the Digital Filterbank Mark III backend \citep[DFB;][]{prandoni_2017_aap}, in {\tt psrfits} format \citep{hotan_2004_pasa}. These data have time and frequency resolutions of 128\,$\upmu$s and 1\,MHz, respectively and a frequency range of $1140.5$--$2163.5$\,MHz, of which $1210.5$--$1739.5$\,MHz is usable (given the receiver response and radio frequency interference).

\section{Analysis and Results}

\subsection{Burst discovery}

For both Effelsberg and Sardinia, the raw voltage data and pulsar backend data were independently searched for bursts using distinct analysis pipelines. We converted the raw voltage data to filterbank format with a time and frequency resolution of 128$\,\upmu $s and 125\,kHz, respectively, using {\tt digifil} \citep{vanstraten_2011_pasa}. We used a Heimdall-based search to identify candidate FRBs, which were then classified using the machine learning classifier FETCH \citep[specifically, models A and H with a 50\% probability threshold;][]{agarwal_2020_mnras}. In this search, we detected 18 bursts in total from \frb, 13 on April 10 (Epoch\,1) and a further 5 on April 19 (Epoch\,2). 

A single-pulse search of the PSRIX and DFB data was conducted using PRESTO tools \citep{ransom_2001_phdt}, including masking of radio frequency interference (RFI) using the {\tt rfifind} tool. The candidate FRBs were then classified using an automated clustering classifier based on \citet{michilli_2018_mnras}. The data quality and analysis strategy was tested using a scan of the test pulsar PSR~J2257+5909. We independently found all 5 bursts from Epoch\,2 in the PSRIX data while only B14 was found in this way in the DFB data from Sardinia. The DFB data were additionally searched using a Spandak-based pipeline\footnote{\url{https://github.com/gajjarv/PulsarSearch}} \citep{Gajjar_2021_AJ, gajjar_2018_apj} which discovered B15 and B16. Post-search analysis revealed that the other bursts were either below the S/N threshold used in the search, or coincident with severe RFI and consequently ruled non-astrophysical by the classifier.

For the remainder of this Letter, the bursts are named B$n$, where $n$ is an integer from 1 to 18, ordered according to the burst arrival time.

\subsection{Localisation}

The interferometric data were correlated using the software correlator SFXC \citep{keimpema_2015_exa} at the Joint Institute for VLBI ERIC (JIVE; the Netherlands), with an integration time of 2\,s and 64 channels per 16\,MHz subband (under the EVN correlation-only proposal EK048, PI: Kirsten). Using the arrival times of the 18 \frb\ bursts detected during our PRECISE campaign, a second correlation was performed for only the data containing bursts, where the gate width used for correlation was determined by eye to maximise the signal-to-noise ratio (S/N). These values are recorded in Table\,\ref{tab:burst_vlbi}. The phase centre used for correlation was the VLA \frb\ localisation position \citep{law_2021_atel}, which has an uncertainty of approximately 1\,arcsecond. This meant that we were not required to correlate the data for a third time to move the phase centre closer to the FRB position, as has been required in previous VLBI FRB projects \citep{marcote_2020_natur,kirsten_2021_arxiv}. 

The analysis steps for interferometric calibration and imaging were performed using standard tasks in the Astronomical Image Processing System, {\tt AIPS} \citep{greisen_2003_assl} and {\tt DIFMAP} \citep{shepherd_1994_baas}. Initially, using the products from the automated EVN Pipeline\footnote{\url{https://evlbi.org/evn-data-access}}, we performed {\it a-priori} amplitude calibration using the gain curves and system temperature measurements from each station. We also applied a bandpass calibration, in addition to some basic flagging when telescopes were off-source. Burst B5 occurred while Effelsberg was still slewing, and so these data were flagged by this initial flagging step. Therefore, we removed these flags from the burst interferometric data to recover the data for burst B5. At Epoch\,1 (EVN correlation project code: EK048D, PRECISE observing code: pr153a), no fringes were detected from Irbene until 17:35~UT, and therefore there is no data from Irbene for bursts B1--B4. The station Badary had reduced time during Epoch\,2 (EK048E, pr156a) and therefore did not observe after 15:45~UT, resulting in no data for this station during any of the bursts detected at Epoch\,2. Delays due to dispersion in the ionosphere are corrected for using electron content maps provided by CDDIS \citep{noll_2010_asr}. We first remove the instrumental delay, i.e. phase jumps between the subbands, using the bright calibrator sources J1048$+$7143 and J0555$+$3948 for Epoch\,1 and 2, respectively. Then we correct the phases for the entire observation, as a function of time and frequency, by performing a fringe-fit using all calibrator sources. Throughout, we use the most sensitive dish, the 100-m Effelsberg telescope, as our reference antenna for calibration. During Epoch\,2, the phase solutions for Irbene were rapidly varying, thus we conclude that the solutions applied to the target are unreliable and we therefore flag these data. In Table\,\ref{tab:burst_vlbi}, we report the ``effective" number of baselines.  This accounts for: the stations that did not record data at the time of the burst; stations that were flagged during calibration; and, additionally, the loss of sensitivity due to the spectra of some bursts peaking in the lower part of the observing band of Effelsberg, where some telescopes did not observe (see Table\,\ref{tab:tele} for frequency coverage per participating telescope, and Figure\,\ref{fig:bursts} for the burst dynamic spectra). 

The phase calibrator, J0502$+$2516, was then imaged using {\tt DIFMAP}, independently for both epochs. During Epoch\,1, the recovered flux from the phase calibrator was 0.179\,Jy, while during Epoch\,2, the flux had dropped to 0.102\,Jy, and the clean image exhibited strong side-lobes, which we attribute to persistent amplitude errors after calibration. Additionally, radio maps of J0502$+$2154 from December 2018 at 5\,GHz and 8.6\,GHz exhibit an unresolved flux of $\sim 0.18$\,Jy\footnote{\url{http://astrogeo.org/cgi-bin/imdb\_get\_source.csh?source\_name=J0502\%2B2516}}, allowing us to assume our higher measured flux is the true value. We therefore use a model of J0502$+$2516 determined by the data from Epoch\,1 to self-calibrate the data from both epochs. This is an appropriate step, since the active galactic nucleus (AGN) J0502$+$2516 is not expected to vary in brightness significantly on $\sim$\,week timescales. The self-calibrated peak brightness of the phase calibrator is 0.178\,Jy\,beam$^{-1}$ and 0.179\,Jy\,beam$^{-1}$ for Epoch\,1 and 2, respectively. The self-calibration solutions were then transferred to the target, before imaging.     

In Figure\,\ref{fig:comp} (left), we plot the clean 1.4-GHz map of all 18 bursts presented in this work. The clean map is produced by fitting a Gaussian component to the visibilities in the {\it uv}-plane, and inverse Fourier transforming using {\tt DIFMAP}'s {\tt modelfit} tool. The combined visibilities from all bursts provides us with the best available {\it uv}-coverage, and so from this we derive the J2000 position of \frb in the International Celestial Reference Frame (ICRF): $\mathrm{RA} = 05^{\rm h}08^{\rm m}03.5073^{\rm s}\pm4.5$\,mas, $\mathrm{Dec} = +26^\circ03^\prime38.5049^{\prime\prime}\pm3.6$\,mas. The peak emission is 17\,$\sigma$, with the next-highest side-lobes measured at approximately 30\,\% of the peak value. Additionally, we applied the same outlined calibration steps to the check source J0501$+$2530, and recovered the source with positional offset $\Delta$RA\,$=1.5$\,mas, $\Delta$Dec\,$=0.6$\,mas from the expected position (as quoted in the rfc2021c catalogue\footnote{http://astrogeo.org/sol/rfc/rfc\_2021c/}). The localisation of \frb\ is therefore robust. The uncertainties on the final position of \frb\ are the quadrature sum of the statistical errors derived from the shape and size of the synthesised beam, normalised by the square root of the S/N ($\Delta$RA$=3.7$\,mas, $\Delta$Dec$=2.5$\,mas), the statistical uncertainty on the phase calibrator position ($\pm0.13$\,mas), and an estimate of the systematic uncertainty due to the separation of the phase calibrator and target (conservatively $\pm$2.5\,mas; \citealt{kirsten_2015_aa}). The position of \frb\ reported here is in agreement (within 2-$\sigma$) of the original EVN localisation reported in \citet{marcote_2021_atel} (which was determined using only the burst data from Epoch 1). Our position for \frb\ is also in agreement with the independent measurements of the VLA \citep{ravi_2021_arxiv}, the Australian Square Kilometre Array Pathfinder (ASKAP) \citep{fong_2021_apjl}, and the upgraded Giant Meterwave Radio Telescope (uGMRT; if one includes the estimated systematic errors they discuss; \citealt{wharton_2021_atel}). These studies are approximately 3 orders of magnitude less precise than the localisation presented here (a comparison is shown in Figure\,\ref{fig:comp}).  

Additionally, we produced the dirty map of the entire $\sim7$ hours of target data from both epochs, to search for any compact persistent radio emission at the site of the FRB, or nearby (Figure\,\ref{fig:comp}; right). The rms noise in the continuum map, using a natural weighting scheme (using {\tt uvweight} in {\tt DIFMAP}) is 14\,$\upmu$Jy\,beam$^{-1}$, and using a uniform weighting scheme is 25\,$\upmu$Jy\,beam$^{-1}$. In addition, we tried applying different Gaussian tapers, 1\,M$\lambda$, 2\,M$\lambda$ and 5\,M$\lambda$ (using {\tt uvtaper} in {\tt DIFMAP}), to downweight the longer baselines and hence boosting any possible extended emission. We find no significant persistent radio source above 6\,$\sigma$, in an area of 8$\times$8\,arcseconds around the FRB. This is in agreement with the original report in \citet{marcote_2021_atel}, and follow-up efforts with the Very Long Baseline Array (VLBA; \citealt{ravi_2021_arxiv}). The shortest baseline in our array is $\sim 270$\,km, between Effelsberg and Westerbork, and thus emission above an angular scale of $\sim 160$\,mas is resolved out. 

\subsection{Burst characterisation}

\begin{figure*}
\resizebox{\hsize}{!}
        {\includegraphics[trim=0cm 0cm 0cm 0cm, clip=true]{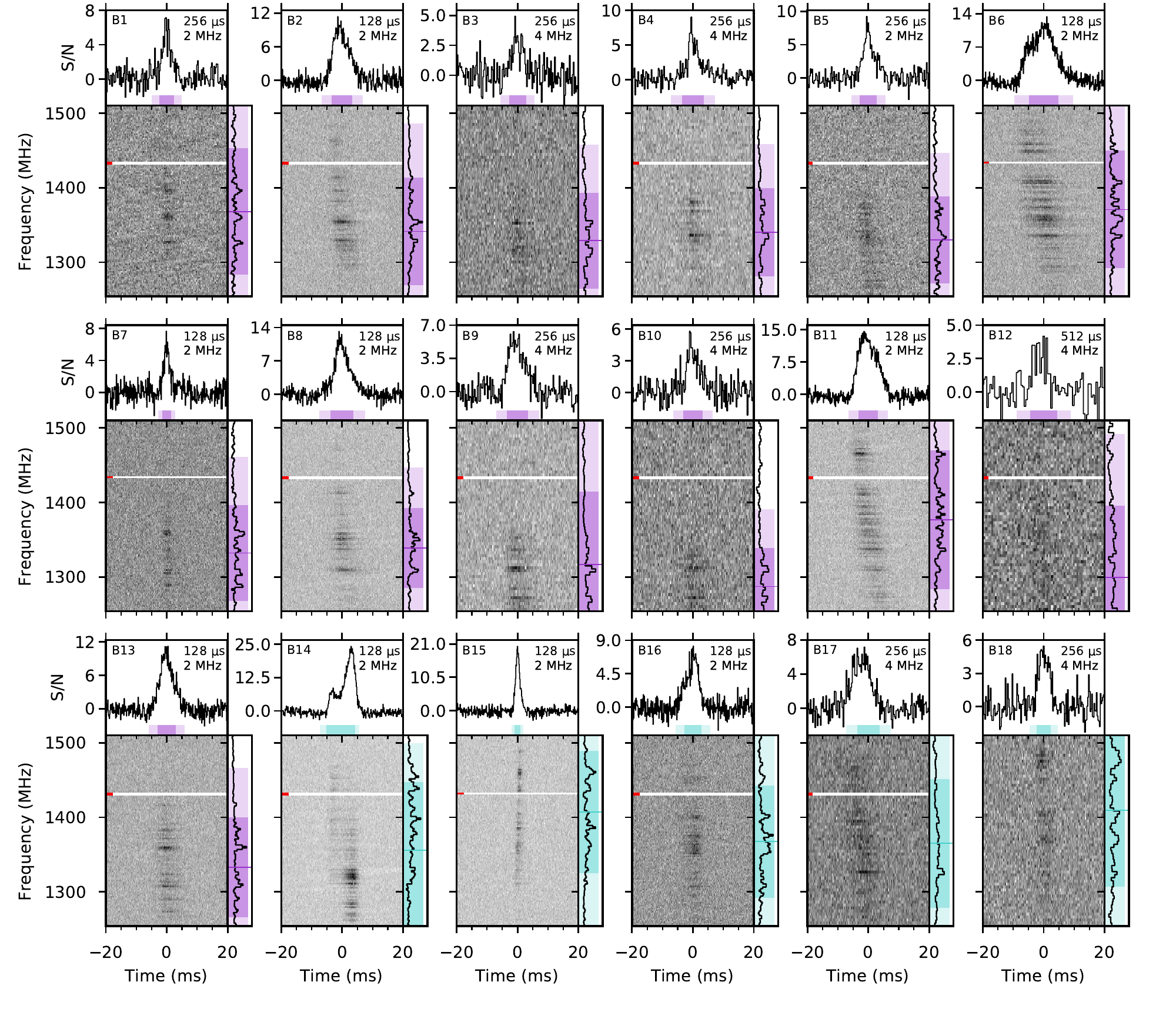}}
  \caption{Dynamic spectra, temporal profiles, and time-averaged spectra for all 18 bursts presented in this work. For each sub-figure, the burst name and time/frequency resolution is shown in the top left and top right, respectively. The coloured bars represent the 1-$\sigma$ (dark) and 2-$\sigma$ (light) error regions of the temporal width and spectral extent of each burst. The purple colour is used for bursts detected during Epoch\,1, and cyan for those detected during Epoch\,2. Data which has been masked due to radio frequency interference is not plotted, and indicated with the red ticks. Each burst has been de-dispersed using a dispersion measure of 412\,pc\,cm$^{-3}$.}
     \label{fig:bursts}
\end{figure*}

We created 32-bit total intensity filterbank data for each burst by autocorrelating the single-dish Effelsberg raw voltage data using SFXC.  These data have time and frequency resolutions of 8\,$\upmu$s and 125\,kHz, respectively.  We then created archive files using {\tt PSRCHIVE} \citep{hotan_2004_pasa}, which have the same resolution as the filterbank data. The dynamic spectra, time profiles, and time-averaged spectra for all bursts are shown in Figure\,\ref{fig:bursts}. The data were coherently dedispersed to a dispersion measure (DM) of 413\,pc\,cm$^{-3}$ \citep{chime_2021_atel}. We optimised the structure (using {\tt DM\textunderscore phase}\footnote{\url{https://github.com/danielemichilli/DM\_phase.git}}) of burst B14, which exhibits clear burst structure, to measure a DM of $412.2\,\pm\,0.6$\,pc\,cm$^{-3}$. Additionally, using B15, which does not exhibit clear structure but is the narrowest burst in our sample, we measure a DM of $411.6\,\pm\,0.4$\,pc\,cm$^{-3}$. Note that for B15, using {\tt DM\textunderscore phase} essentially maximises the S/N since it does not exhibit clear burst structure. The final DM, which we used to incoherently shift the frequency channels for all bursts in our sample, assuming the DM does not vary significantly on $<10$\,day timescales, was determined by averaging the two measurements: $\mathrm{DM} = 412.0\,\pm\,0.7$\,pc\,cm$^{-3}$. 

To measure the temporal and spectral extent of the bursts, we performed a two-dimensional autocorrelation of each dynamic spectrum. The time of arrival and central frequency of the bursts were determined using a two-dimensional Gaussian fit to the dynamic spectrum. This method of peak and width determination is explained in more detail in \citet{nimmo_2021_arxiv}, and the values for the bursts in this work are reported in Table\,\ref{tab:burst_properties}. 

Within the 2\,$\sigma$ burst width and spectral extent, indicated by the light purple and light cyan bars in Figure\,\ref{fig:bursts}, we compute the burst fluence using the radiometer equation \citep{cordes_2003_apj}. For this, we use the typical Effelsberg system temperature, 20\,K, combined with a cosmic microwave background contribution of 3\,K, and a sky background temperature of 1\,K, which we obtain by extrapolating from the 408\,MHz sky map \citep{remazeilles_2015_mnras} using a spectral index of $-$2.7 \citep{reich_1988_aa}. We also use the typical Effelsberg gain of 1.54\,Jy/K. The typical system values for Effelsberg are uncertain on the 20\,\% level, dominating the error on the fluence and peak flux density measurements reported in Table\,\ref{tab:burst_properties}. 

We computed the one-dimensional auto-correlation function (ACF) of the burst spectra in order to measure the scintillation bandwidth. The scintillation bandwidth is defined as the half-width at half-maximum (HWHM) of a Lorentzian fit to the ACF of the spectrum. We note that the zero-lag noise spike is removed from each ACF. Additionally, we subtract the off-burst ACF to reduce the power at low lags due to noise. The narrow-bandedness of many of the bursts in our sample results in an additional frequency structure in the ACF with a characteristic bandwidth related to that of the frequency extent of the burst. For all bursts, other than B1, B15, B17 and B18, this broadband feature is visible in the ACF. We fit a one-dimensional Gaussian function to this wider component and subtract it from the ACF in order to disentangle the two frequency scales. Finally, we fit a Lorentzian function to the remaining structure in the ACF, and measure the scintillation bandwidth per burst (reported in Table\,\ref{tab:burst_properties}), which we find to be consistent with the results presented in \citet{main_2021_arxiv}.

\section{Discussion}

\subsection{Astrometry}

\begin{figure*}
\resizebox{\hsize}{!}
        {\includegraphics[trim=0cm 0cm 0cm 0cm, clip=true]{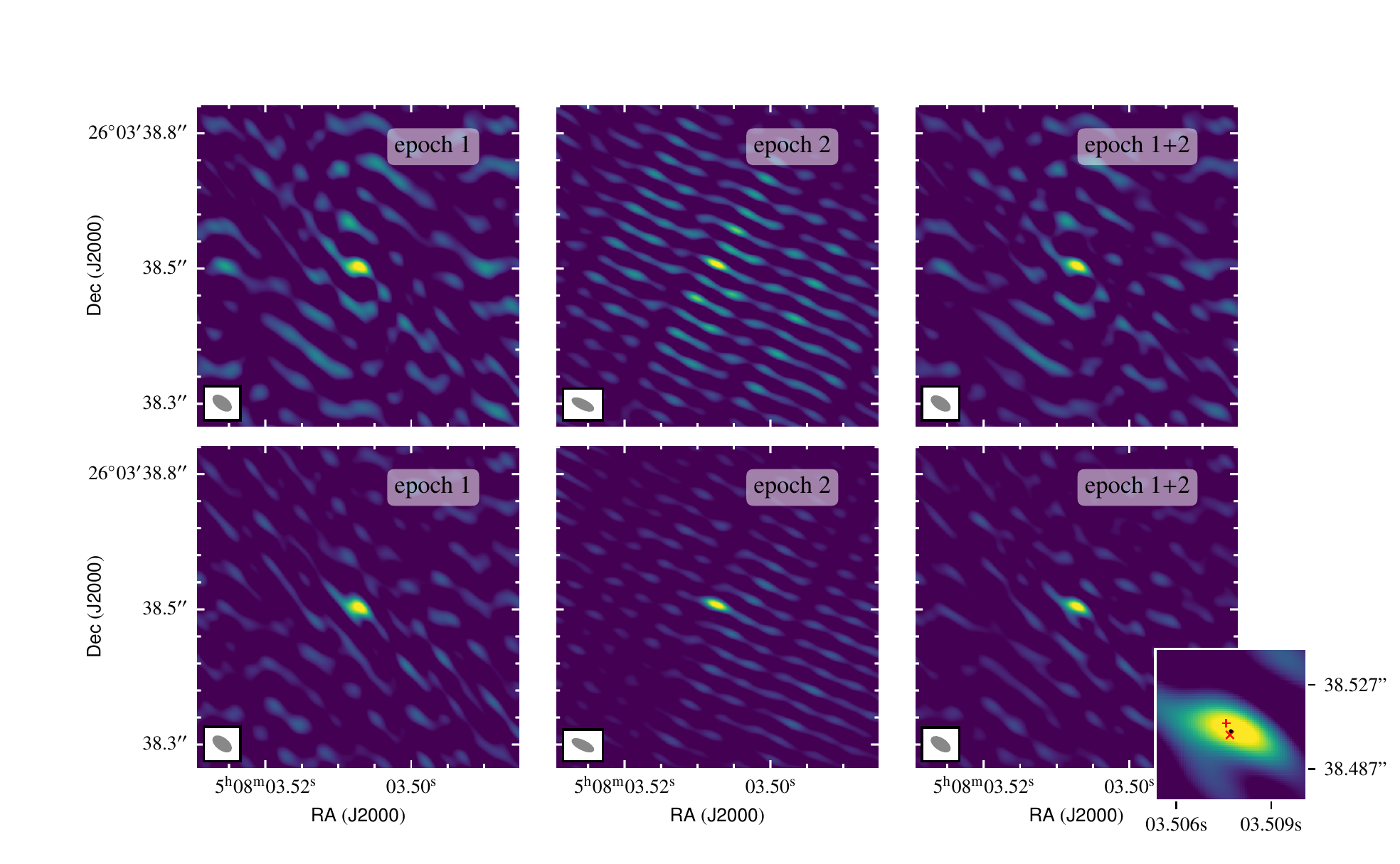}}
  \caption{Top row: dirty maps (i.e. the Fourier transform of the visibilities) of the combined visibilities of the 13 bursts discovered during Epoch\,1 (left), the 5 burst during Epoch\,2 (middle) and all bursts from both epochs (right). Bottom row: same as the top row except showing clean maps (i.e. the Fourier transform of the visibilities after deconvolving the telescope response). The zoom-in panel on the bottom right sub-plot shows the best-fit two-epoch position in black, with the positions from Epoch\,1 and 2 independently represented by the red cross and plus, respectively. All images are made using a natural weighting scheme, and the scaling of each radio map goes from 5\,\% to 85\,\% of the peak value. The synthesised beam is shown at the bottom left of each panel. }
     \label{fig:localisation}
\end{figure*}

The final position of \frb\ is determined from a combined image of all $18$ bursts presented in this work.  The combination of many bursts, at two epochs, maximises the {\it uv}-coverage.  We also investigate the position of \frb\ per observing epoch, in order to explore how consistent the FRB position is with an independent calibration, and observed on separate days. The position of \frb\ using the combined visibilities of 13 bursts at Epoch\,1 is $\mathrm{RA}= 05^{\rm h}08^{\rm m}03.5073^{\rm s}\pm4.7$\,mas, $\mathrm{Dec}=+26^\circ03^\prime38.5032^{\prime\prime}\pm3.9$\,mas (following the error determination described in Section\,3.2, and using the Epoch 1 beam shape and S/N of 13.8). Similarly, the position of \frb\ using the combined visibilities of 5 bursts at Epoch\,2 is $\mathrm{RA}=05^{\rm h}08^{\rm m}03.5074^{\rm s}\pm5.9$\,mas, $\mathrm{Dec}=+26^\circ03^\prime38.5087^{\prime\prime}\pm3.5$\,mas (Epoch 2 S/N 11.9). Both per-epoch positions agree with the combined-epoch final position, and with each other, within 2-$\sigma$. In Figure\,\ref{fig:localisation}, we plot the dirty maps (i.e. the inverse Fourier transform of the visibilities) of the combined visibilities from all 13 bursts in Epoch\,1, the combined visibilities from all five bursts in Epoch\,2, and the combined visibilities of all bursts from both epochs. Also shown in Figure\,\ref{fig:localisation} are the corresponding clean maps, with a visual comparison of the astrometry per epoch.  

\begin{figure*}[!]
\resizebox{\hsize}{!}
        {\includegraphics[trim=0cm 0cm 0cm 0cm, clip=true]{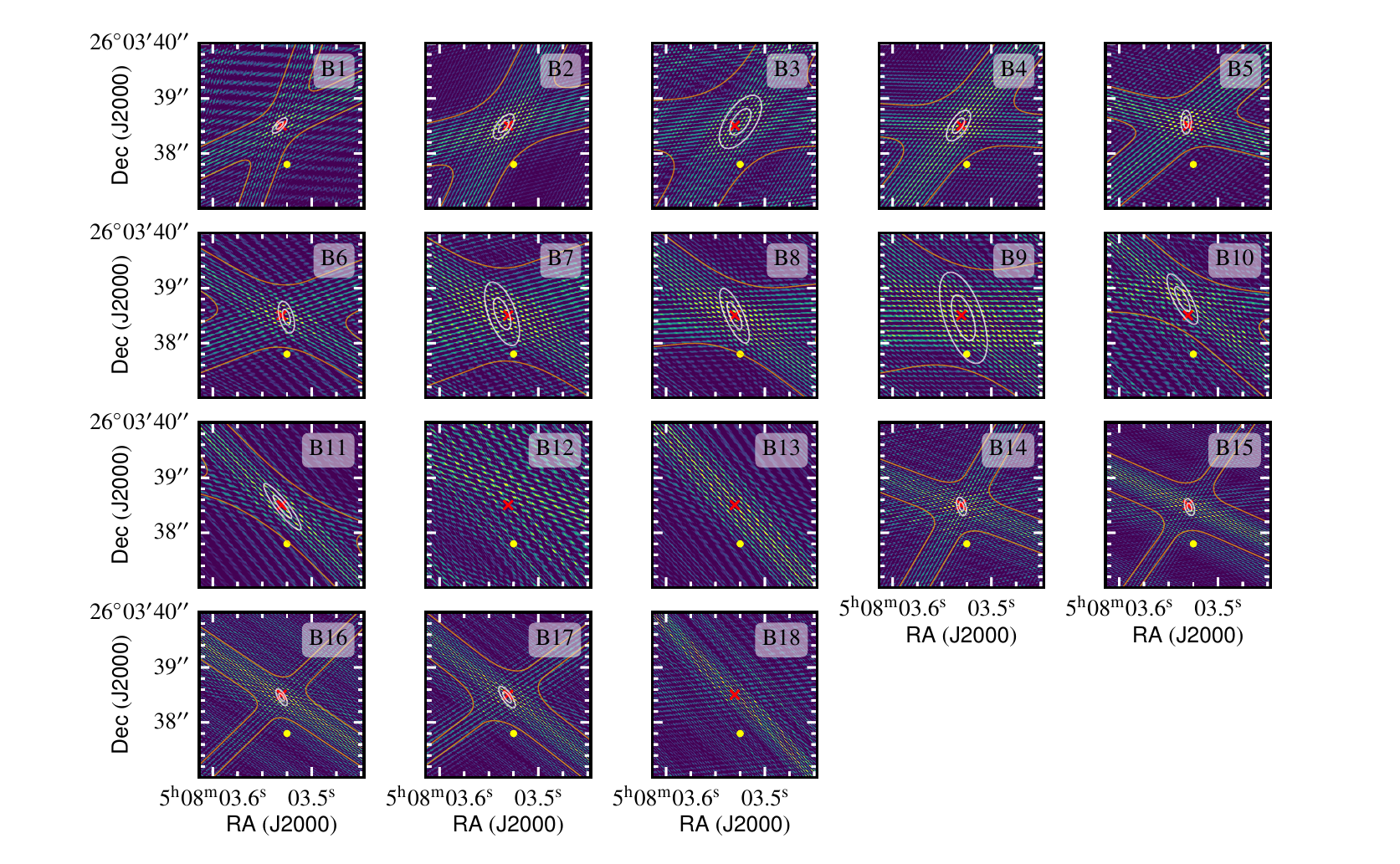}}
  \caption{Dirty maps of the individual bursts, with the burst name shown at the top right corner of each panel. The orange contours represent the two 2-dimensional Gaussian fit (1-$\sigma$ region) to the cross fringe pattern. The white contours represent the 1-$\sigma$ and 2-$\sigma$ Gaussian fit to the intersection of the double Gaussian, normalised by the S/N of the double Gaussian peak. Note: B12, B13 and B18 do not have a constraining Gaussian fit due to the lack of cross pattern in the dirty map. The best-fit \frb\ position using all 18 bursts is shown by the red cross on each panel, and the phase centre used for correlation is indicated by the yellow point.}
     \label{fig:dirty}
\end{figure*}

\begin{figure*}
\resizebox{\hsize}{!}
        {\includegraphics[trim=0.2cm 0.3cm 0.3cm 1.2cm, clip=true]{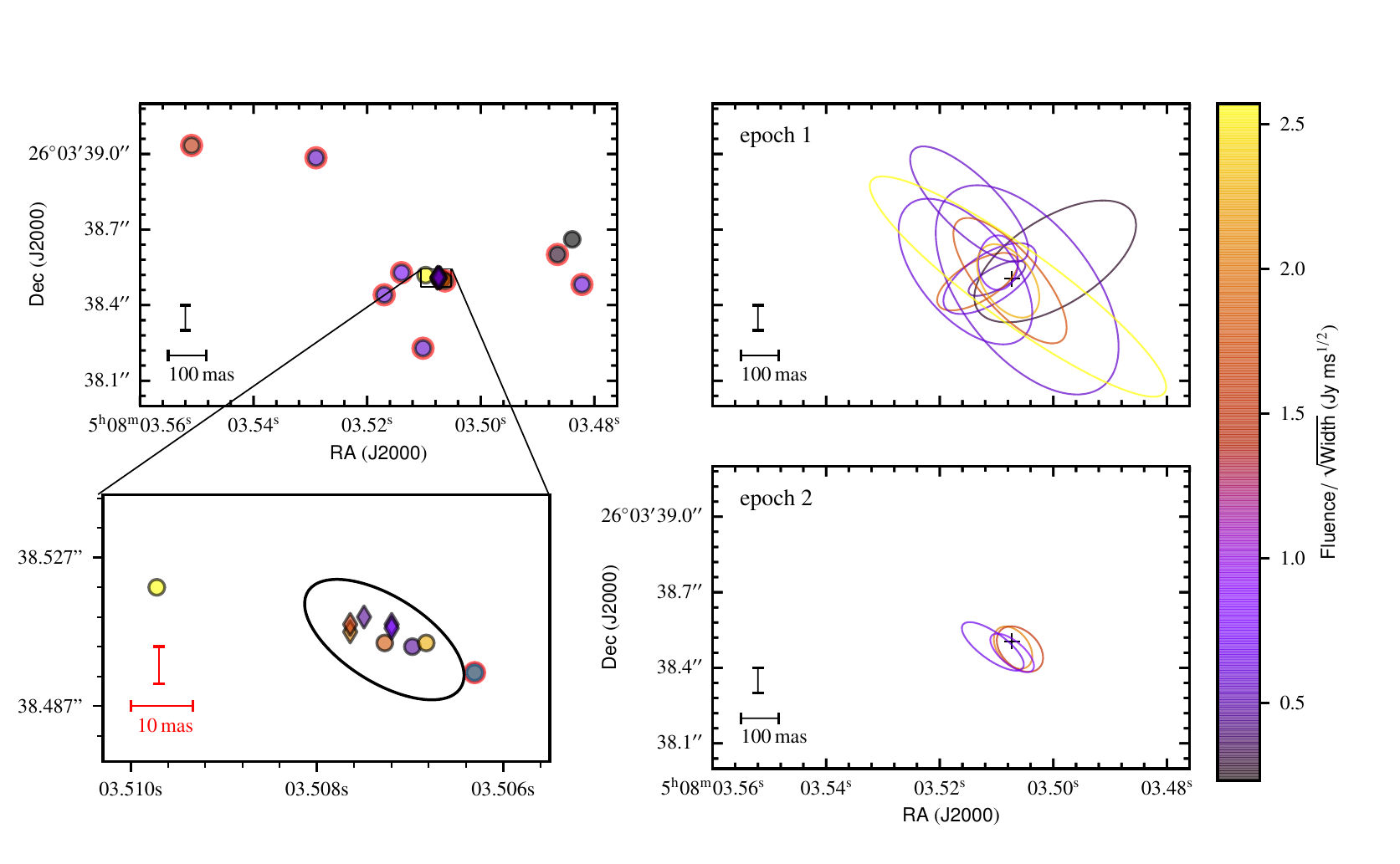}}
  \caption{Individual burst positions determined using the intensity peak on the dirty maps (left) and a Gaussian fit to the dirty maps which exhibit a clear cross fringe pattern (right; see also Figure\,\ref{fig:dirty}). Circles indicate bursts from Epoch 1 and diamonds bursts from Epoch 2. The colourbar represents the burst fluence divided by the square-root of the burst width, and is used as a measure of the burst detectability (a proxy for S/N; \citealt{marcote_2017_apjl}). The black bars represent a 100\,mas angular scale, while the red bars in the zoom-in panel represent 10\,mas angular scale. The black ellipse on the left bottom plot represents the synthesised beam centred on the best-fit \frb\ position, while the cross on the  right plots represent the best-fit \frb\ position (in both cases, using all 18 bursts). The  points on the left plot outlined in red indicate the bursts that were detected with only $6$ baselines, highlighting the impact of low number of baselines on the scatter of individual burst positions. }
     \label{fig:bursts_pos}
\end{figure*}

To explore the astrometry further, we determine the positions of each burst individually, using both the peak positions on the dirty map (dominated by the long baselines in the array), and a Gaussian fit to the fringe pattern (shown in Figure\,\ref{fig:dirty}; dominated by the short baselines). For this Gaussian method, we first fit 2-dimensional Gaussians to both arms of the cross pattern in the fringes, and then fit a single 2-dimensional Gaussian to their intersection. Note that we include the Gaussian fits only for bursts that exhibit a clear cross fringe pattern in their dirty map (Figure\,\ref{fig:dirty}). A Gaussian is not necessarily the correct function to fit to the dirty map (the true function is dependent on the {\it uv}-coverage). However, it is clear from the two two-dimensional Gaussian fits in Figure\,\ref{fig:dirty}, and the fact that $>68$\,\% of the 1-$\sigma$ Gaussians in Figure\,\ref{fig:bursts_pos} intersect with the position of \frb, that this is a conservative approach to measure the intersection of the cross fringe pattern. In future work, we will also use these data to test a more sophisticated method in which we simultaneously fit the fringes of all baselines, using the known burst amplitudes, baseline lengths, and orientations.  In such a scheme, we would fit only for the unknown burst position, and take maximum advantage of the known signal and instrument parameters.  Such a method could potentially increase the precision and accuracy of positional determinations in the case where only one burst is detected and/or the number of effective baselines is low.

\begin{figure*}
\resizebox{\hsize}{!}
        {\includegraphics[trim=0.3cm 0.3cm 0.3cm 1.2cm, clip=true]{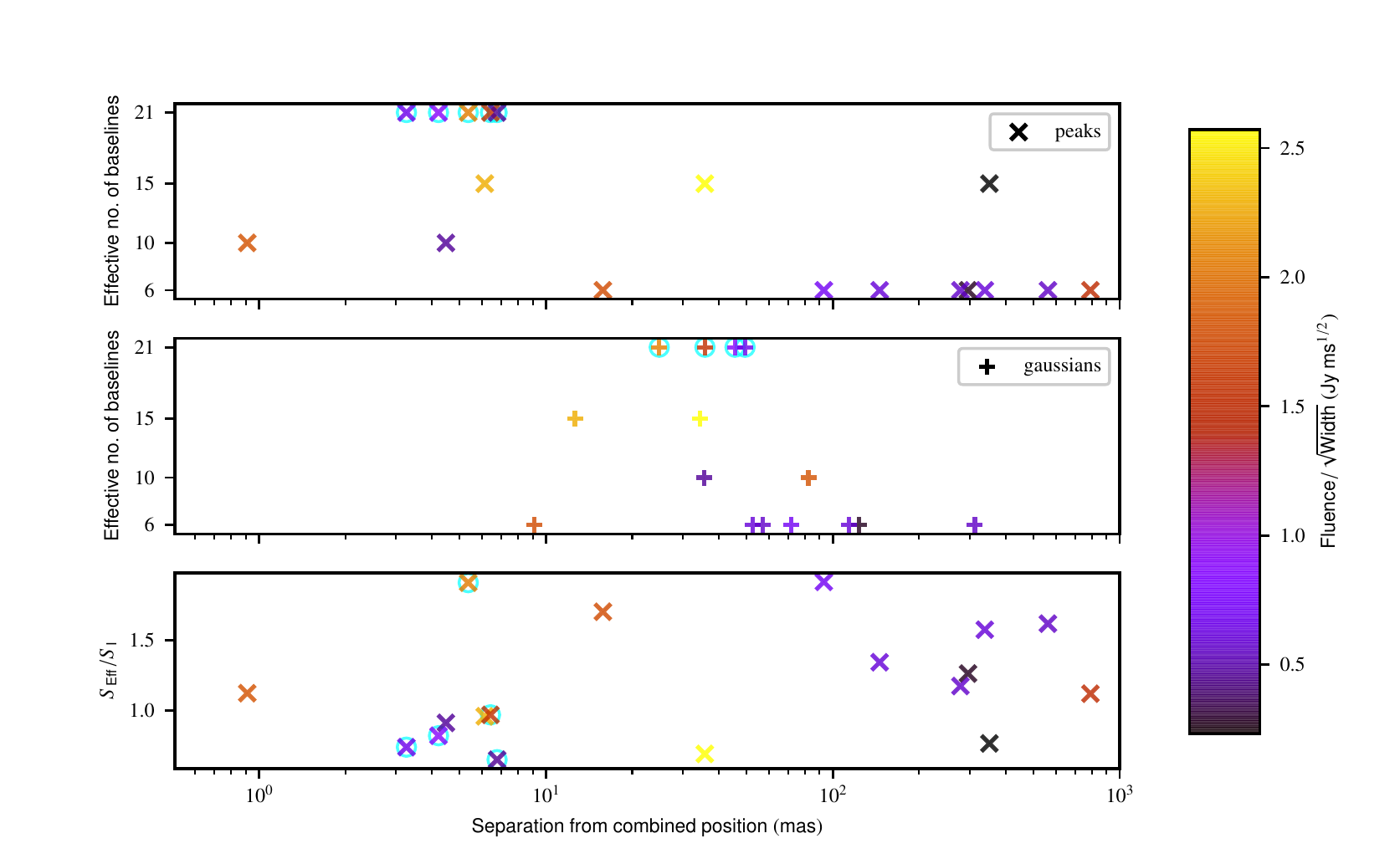}}
  \caption{Separation of the peak (top) and Gaussian fit (middle) positions of the individual bursts as presented in Figure\,\ref{fig:bursts_pos}, from the best-fit \frb\ position (using all 18 bursts) --- as a function of effective number of baselines (defined in the text), and the detectability parameter, ${\rm Fluence}/\sqrt{\rm Width}$. The bottom panel shows the separation of the peak individual burst positions from the best-fit \frb\ position, as a function of the burst peak flux density measured in the Effelsberg single-dish data ($S_{\sf Eff}$) divided by that measured in the interferometric data ($S_{\sf I}$). On all panels, the markers with the cyan circles indicate the bursts from Epoch 2.  }
     \label{fig:bursts_sep}
\end{figure*}

The positions using both the peak and Gaussian fit methods are reported in Table\,\ref{tab:burst_vlbi}, and illustrated in Figure\,\ref{fig:bursts_pos}. In Figure\,\ref{fig:bursts_sep}, we show the separation of the peak and Gaussian positions per burst, from the best-fit position of \frb. As a result of the larger number of baselines ($=N(N-1)/2=21$, for number of telescopes $N$) during the second epoch, compared with the first, the peak positions of the Epoch\,2 bursts agree within 10\,mas of the true position. For the bursts from Epoch\,1 (for which we have a lower number of baselines), the scatter in the peak positions increases and deviates farther from the true position, as expected. Accurately determining the astrometric error on the individual peak positions in the regime of low number of baselines, would likely require an empirical statistical study similar to \citet{martividal_2010_aa}. However, based on Figure\,\ref{fig:bursts_sep}, an arcsecond localisation is feasible given one burst above a detectability threshold of Fluence/$\sqrt{\rm Width} \approx 0.3$\,Jy\,ms$^{1/2}$ \citep{marcote_2017_apjl}, detected with only six baselines (four telescopes).

By looking at the phases of the visibilities of the individual bursts, there is clear scatter between approximately 50$^\circ$ and 100$^\circ$. This scatter arises due to a combination of phase noise, and errors in phase-referencing. The phase error from phase noise is related to the S/N approximately as 1/(S/N). Therefore, for a S/N of 3, one can expect a phase error of approximately 20$^\circ$. The uncertainty on the phase referencing accuracy arises due to a combination of separation of phase calibrator and target, atmospheric conditions, and accuracy of the correlation model. \citet{martividal_2010_aa_2, martividal_2010_aa} derive expressions to estimate VLBI astrometric errors depending on the separation of phase calibrator and target, observing frequency, integration time, and the telescope's diffraction limit.  Although this approach allows for the negligible integration time for individual bursts, the small number of baselines will have a significant effect on the astrometric accuracy in this limiting case. Therefore, the expressions for astrometric accuracy derived in \citet{martividal_2010_aa_2,martividal_2010_aa} will underestimate the errors on the astrometry of individual bursts in this work, due to the lack of {\it uv}-coverage. Assuming we have a realistic phase error, in our sparse {\it uv}-coverage example, of $100^\circ$, arising from a combination of phase noise and calibration uncertainty, the expected positional shift of the fringe pattern is $\sim 1.22\lambda/b \times 100/180 \approx 112$\,mas, where $\lambda$ is our central observing wavelength, and $b$ in this case is the shortest baseline of 270\,km between Effelsberg and Westerbork (shortest baseline is relevant here since the cross fringe pattern is dominated by the short baselines). This is consistent with the scatter evident in the Gaussian fit positions on Figure\,\ref{fig:bursts_sep} (mean\,$=71$\,mas, with standard deviation\,$=72$\,mas). It is reasonable to expect an even larger scatter on the peak positions since this combines the errors on the short baselines, as well as on the long baselines. Additionally, the peak positions will be highly influenced by the phase noise on long baselines, which in general is larger than on shorter baselines. We indeed observe a larger mean and scatter on the peak positions than on the Gaussian fit positions (Figure\,\ref{fig:bursts_sep}; peaks mean\,$=164$\,mas, with standard deviation\,$=223$\,mas).

Therefore, due to the lower scatter in the Gaussian-fit positions in Figure\,\ref{fig:bursts_sep} compared with the peak positions, we conclude that the safer approach to determine individual burst positions, with an array of $<20$\,baselines, is by using the Gaussian fit method. In this work, we are assuming that ionospheric turbulence is not varying drastically between observing epochs, or at least that our calibration is correcting for this sufficiently accurately: a fair assumption due to our relatively high observing frequency ($>1$\,GHz). Here we have provided an empirical investigation of astrometric uncertainties using one source at two epochs. This does not allow us to consider a wide range of observing conditions, so future such studies on other sources will be useful in our understanding of the limitations of VLBI FRB work.  

\subsection{The local environment of \frb}

\begin{figure*}
\resizebox{\hsize}{!}
        {\includegraphics[trim=0cm 0cm 0cm 0cm, clip=true]{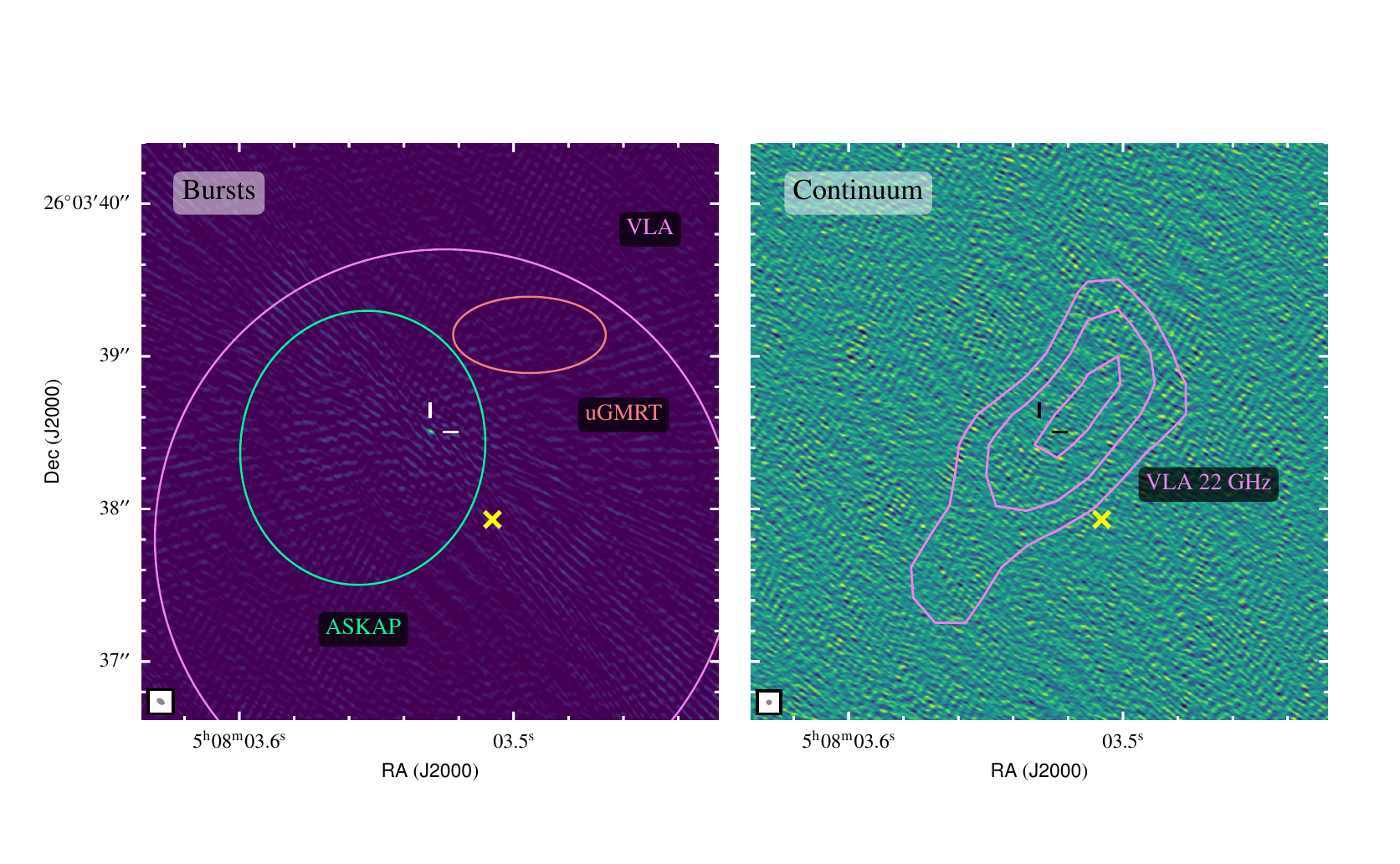}}
  \caption{Left panel: Clean EVN 1.4-GHz map of the combined visibilities of all 18 bursts detected during both epochs (i.e. the Fourier transform of the visibilities after deconvolving the telescope response). The white ticks indicate the position of \frb. The localisation regions of \frb\ as reported by VLA \citep{ravi_2021_arxiv}, ASKAP \citep{fong_2021_apjl} and uGMRT \citep{wharton_2021_atel} are overplotted using pink, green and orange lines, respectively. Right panel: Dirty EVN map (before deconvolution of the interferometer response) of all continuum target data from both epochs to search for persistent radio emission. The position of \frb\ is indicated by the black ticks. Overplotted are the 3, 4 and 5-$\sigma$ contours of the resolved radio emission detected by the VLA at 22\,GHz \citep{piro_2021_arxiv}. In both panels, the optical centre of the host galaxy is indicated by the yellow cross \citep{fong_2021_apjl}. The synthesised beam is shown at the bottom left of each panel. Both maps are made using a natural weighting scheme.}
     \label{fig:comp}
\end{figure*}

The host galaxy of \frb, SDSS J050803.48+260338.0 or hereafter J0508+2603 \citep{day_2021_atel} is a massive, star-forming galaxy \citep{ravi_2021_arxiv,fong_2021_apjl} at a redshift of $z=0.098$ \citep{kilpatrick_2021_atel}. The VLA (D-configuration; \citealt{ricci_2021_atel}) detected unresolved compact persistent emission at 3\,GHz and 9\,GHz, in addition to the uGMRT \citep{wharton_2021_atel_prs} detection of unresolved persistent emission at 600\,MHz. Follow-up with 22-GHz VLA observations in C-configuration, allowed for the emission to be resolved \citep{piro_2021_arxiv}. The lack of compact emission in our EVN 1.4-GHz observations (Figure\,\ref{fig:comp}), supports the conclusion that the radio emission seen with lower resolution instruments is from star-formation \citep{ravi_2021_arxiv,fong_2021_apjl,piro_2021_arxiv}. 

The milliarcsecond precision of our EVN localisation, allows us to explore where the FRB location is relative to the radio star-formation emission \citep{piro_2021_arxiv}, and the centre of the host galaxy \citep{fong_2021_apjl}. We find that \frb\ is 705\,$\pm$26\,mas (projected distance: $\sim$1.3\,kpc, assuming an angular size distance of 375.9\,Mpc; \citealt{kilpatrick_2021_atel}), from the optical centre of the host galaxy, statistically inconsistent with originating from the galaxy centre, similar to the discussion in \citet{fong_2021_apjl}. The uncertainties on this offset arise as the quadrature sum of the radio position uncertainty ($4.5$\,mas), the optical position uncertainty in Pan-STARRS (13\,mas; \citealt{fong_2021_apjl}), and the astrometric tie uncertainty between Pan-STARRS and {\em Gaia} (22\,mas; \citealt{magnier_2020_apjs}). We note that the {\em Gaia} reference frame and the ICRF agree on the few-milliarcsecond level \citep{mignard_2016_aa}, therefore the uncertainties on the optical position and frame tying Pan-STARRS to {\em Gaia} dominate the error budget.  \frb\ is $175\pm180$\,mas from the peak of the radio star-formation emission, where the uncertainty is dominated by the positional accuracy of the peak of the extended 22-GHz emission \citep{piro_2021_arxiv}.

Future observations in optical and infrared using high resolution instruments such as the {\it Hubble Space Telescope}, will allow for a measurement of the proximity of \frb\ with star-forming knots in the host galaxy. This can be compared with the measured 250\,pc and 260\,pc offset from the peak of a nearby star-forming region in the case of FRB~20180916B \citep{tendulkar_2021_apjl}, and FRB~20121102A \citep{bassa_2017_apjl,kokubo_2017_apj}, respectively. Additionally, these observations will allow for exploration of the role of star-formation on the period of high activity \citep{lanman_2021_arxiv}, the production of extremely bright bursts \citep{kirsten_2021_atel, herrmann_2021_atel}, as well as the presence of significant circular polarisation and polarisation angle swings in some bursts from \frb\ \citep{hilmarsson_2021_mnras, kumar_2021_arxiv}.

\begin{table*}
\caption{\label{tab:tele}Individual telescope configurations used during our interferometric observations.}
\resizebox{\textwidth}{!}{\begin{tabular}{lccc}
\hline
\hline
		{Telescope} & {Frequency coverage (MHz)} & {No. of subbands} & {Epoch$\mathrm{^{a}}$} \\
		\hline
		{Effelsberg (Ef)} & {1254 -- 1510} & {16\,$\times$\,16\,MHz} & {1,2}\\
		{Onsala (O8)} & {1254 -- 1510} & {16\,$\times$\,16\,MHz} & {1,2}\\
		{Toru\'n (Tr)} & {1254 -- 1510} & {16\,$\times$\,16\,MHz} & {1,2}\\
		{Irbene (Ir)} & {1382 -- 1510} & {8\,$\times$\,16\,MHz} & {1,2}\\
		{Westerbork (Wb)} & {1382 -- 1510} & {8\,$\times$\,16\,MHz} & {1,2}\\
		{Noto (Nt)} & {1318 -- 1574} & {16\,$\times$\,16\,MHz} & {1}\\
		{Medicina (Mc)} & {1350 -- 1478} & {8\,$\times$\,16\,MHz} & {2}\\
		{Svetloe (Sv)} & {1382 --  1510} & {8\,$\times$\,16\,MHz} & {2}\\
		{Badary (Bd)} & {1382 -- 1510} & {8\,$\times$\,16\,MHz} & {2}\\
		{Sardinia (Sr)} & {1350 -- 1606} & {8\,$\times$\,32\,MHz} & {2}\\
		\hline
		\multicolumn{4}{l}{$\mathrm{^{a}}$ Epochs which the telescope participated, where Epoch 1 corresponds to EVN project code EK048D,}\\
		\multicolumn{4}{l}{\hspace{0.5em}and Epoch 2 is EK048E.}\\
\end{tabular}}
  \label{tab:tele}      
\end{table*}	

\begin{table*}
\caption{\label{tab:burst_vlbi}Interferometric burst properties.}
\resizebox{\textwidth}{!}{\begin{tabular}{lcccccccc}
\hline
\hline
		{Burst}& {Peak position$^{\rm a}$} & {Gaussian position$^{\rm b}$} & {$\sigma_{\rm maj}^{\rm \text{ }c}$} & {$\sigma_{\rm min}^{\rm \text{ }d}$} & {$\theta \mathrm{^{\text{ } e}}$} & {Peak flux density$\mathrm{^{f}}$} & {Gate width} & {Effective no. of baselines$\mathrm{^{\text{ } g}}$}  \\
		{} & {[RA, Dec (J2000)]} & {[RA, Dec (J2000)]} & {[mas]} & {[mas]} & {[deg]} &{[Jy/beam]} & {[ms]} & {}\\
		\hline
		Epoch 1 \\
		B1 & $05{\rm h}08{\rm m}03.5070{\rm s}$, $+26^\circ03^\prime38.503^{\prime\prime}$  & $05{\rm h}08{\rm m}03.5099{\rm s}$, $+26^\circ03^\prime38.5068^{\prime\prime}$  & 496.1  & 196.5  & 46.4  & 0.34\,$\pm$\,0.06  & 4.12 & 10 \\
		B2 & $05{\rm h}08{\rm m}03.5073{\rm s}$, $+26^\circ03^\prime38.504^{\prime\prime}$ & $05{\rm h}08{\rm m}03.5133{\rm s}$, $+26^\circ03^\prime38.4931^{\prime\prime}$ & 731.9 & 331.5 & 52.5 & 0.65\,$\pm$\,0.05 & 8.49 & 10\\
		B3 & $05{\rm h}08{\rm m}03.4865{\rm s}$, $+26^\circ03^\prime38.601^{\prime\prime}$ & $05{\rm h}08{\rm m}03.4997{\rm s}$, $+26^\circ03^\prime38.5736^{\prime\prime}$ & 1029.9 & 498.5 & 55.8 & 0.19\,$\pm$\,0.04 & 4.02 & 6\\
		B4 & $05{\rm h}08{\rm m}03.5170{\rm s}$, $+26^\circ03^\prime38.441^{\prime\prime}$ & $05{\rm h}08{\rm m}03.5112{\rm s}$, $+26^\circ03^\prime38.5055^{\prime\prime}$ & 773.1 & 353.3 & 55.8 & 0.32\,$\pm$\,0.05 & 5.15 & 6\\
		B5 & $05{\rm h}08{\rm m}03.5140{\rm s}$, $+26^\circ03^\prime38.529^{\prime\prime}$ & $05{\rm h}08{\rm m}03.5097{\rm s}$, $+26^\circ03^\prime38.5686^{\prime\prime}$ & 624.6 & 270.7 & 89.9 & 0.24\,$\pm$\,0.03 & 7.50 & 6\\
		B6 & $05{\rm h}08{\rm m}03.5068{\rm s}$, $+26^\circ03^\prime38.504^{\prime\prime}$ & $05{\rm h}08{\rm m}03.5079{\rm s}$, $+26^\circ03^\prime38.496^{\prime\prime}$ & 762.3 & 324.8 & 105.6 & 0.95\,$\pm$\,0.06 & 14.54 & 15\\
		B7 & $05{\rm h}08{\rm m}03.4822{\rm s}$, $+26^\circ03^\prime38.482^{\prime\prime}$ & $05{\rm h}08{\rm m}03.5155{\rm s}$, $+26^\circ03^\prime38.5333^{\prime\prime}$ & 1203.2 & 487.3 & 110.8 & 0.33\,$\pm$\,0.05 & 3.78 & 6\\
		B8 & $05{\rm h}08{\rm m}03.5063{\rm s}$, $+26^\circ03^\prime38.496^{\prime\prime}$ & $05{\rm h}08{\rm m}03.5076{\rm s}$, $+26^\circ03^\prime38.4971^{\prime\prime}$ & 1267.3 & 382.9 & 114.4 & 0.54\,$\pm$\,0.06 & 10.75 & 6\\
		B9 & $05{\rm h}08{\rm m}03.5102{\rm s}$, $+26^\circ03^\prime38.229^{\prime\prime}$ & $05{\rm h}08{\rm m}03.5046{\rm s}$, $+26^\circ03^\prime38.4606^{\prime\prime}$ & 1535.8 & 611.9 & 109.0 & 0.23\,$\pm$\,0.04 & 8.47 & 6\\
	    B10 & $05{\rm h}08{\rm m}03.5290{\rm s}$, $+26^\circ03^\prime38.985^{\prime\prime}$  & $05{\rm h}08{\rm m}03.5148{\rm s}$, $+26^\circ03^\prime38.8008^{\prime\prime}$ & 1088.9 & 334.2 & 118.2 & 0.21\,$\pm$\,0.03 & 6.01 & 6\\
	    B11 & $05{\rm h}08{\rm m}03.5097{\rm s}$, $+26^\circ03^\prime38.519^{\prime\prime}$ & $05{\rm h}08{\rm m}03.5062{\rm s}$, $+26^\circ03^\prime38.4736^{\prime\prime}$ & 1327.6 & 274.1 & 128.2 & 1.29\,$\pm$\,0.20 & 10.56 & 15\\
	    B12 & $05{\rm h}08{\rm m}03.4839{\rm s}$, $+26^\circ03^\prime38.661^{\prime\prime}$ & -- & -- & -- & -- & 0.17\,$\pm$\,0.03 & 5.12 & 15\\
		B13 & $05{\rm h}08{\rm m}03.5509{\rm s}$, $+26^\circ03^\prime39.034^{\prime\prime}$ & -- & --& -- & --& 0.67\,$\pm$\,0.06 & 8.18 & 6\\
		\hline
		Epoch 2\\
		B14 & $05{\rm h}08{\rm m}03.5076{\rm s}$, $+26^\circ03^\prime38.507^{\prime\prime}$ & $05{\rm h}08{\rm m}03.5071{\rm s}$, $+26^\circ03^\prime38.4802^{\prime\prime}$ & 530.1 & 249.9 & 105.6 & 0.79\,$\pm$\,0.05 & 9.44 & 21 \\
		B15 & $05{\rm h}08{\rm m}03.5076{\rm s}$, $+26^\circ03^\prime38.509^{\prime\prime}$ & $05{\rm h}08{\rm m}03.5059{\rm s}$, $+26^\circ03^\prime38.4744^{\prime\prime}$ & 450.9 & 235.8 & 108.0 & 1.27\,$\pm$\,0.10 & 3.14 & 21 \\
		B16 & $05{\rm h}08{\rm m}03.5072{\rm s}$, $+26^\circ03^\prime38.509^{\prime\prime}$ & $05{\rm h}08{\rm m}03.5072{\rm s}$, $+26^\circ03^\prime38.4593^{\prime\prime}$ & 503.0 & 198.4 & 119.8 & 0.61\,$\pm$\,0.06 & 6.54 & 21 \\
		B17 & $05{\rm h}08{\rm m}03.5072{\rm s}$, $+26^\circ03^\prime38.508^{\prime\prime}$ & $05{\rm h}08{\rm m}03.5106{\rm s}$, $+26^\circ03^\prime38.4852^{\prime\prime}$ & 665.5 & 238.6 & 125.1 & 0.42\,$\pm$\,0.03 & 7.07 & 21 \\
		B18 & $05{\rm h}08{\rm m}03.5075{\rm s}$, $+26^\circ03^\prime38.511^{\prime\prime}$ & -- & -- & -- & -- & 0.37\,$\pm$\,0.04 & 5.52 & 21 \\
		
		\hline
		\multicolumn{9}{l}{$\mathrm{^{a}}$ Position of the intensity peak on the dirty map.}\\
		\multicolumn{9}{l}{$\mathrm{^{b}}$ Centroid of the 2-dimensional Gaussian fit to the intersection of the cross fringe pattern in the dirty map.}\\
		\multicolumn{9}{l}{$\mathrm{^{c}}$ The 1-$\sigma$ major axis of the Gaussian}\\
		\multicolumn{9}{l}{$\mathrm{^{d}}$ The 1-$\sigma$ minor axis of the Gaussian}\\
		\multicolumn{9}{l}{$\mathrm{^{e}}$ The rotational angle of the Gaussian. Measured anti-clockwise.}\\
		\multicolumn{9}{l}{$\mathrm{^{f}}$ Determined using the peak of the individual burst dirty maps.}\\
		\multicolumn{9}{l}{$\mathrm{^{g}}$ As is clear in Figure\,\ref{fig:bursts}, some bursts are narrowband with emission mainly below 1380\,MHz.}\\
		\multicolumn{9}{l}{\hspace{0.3cm}Some of the participating telescopes did not observe the lower frequencies (Table\,\ref{tab:tele}) and therefore have reduced sensitivity to these bursts.}\\
		\multicolumn{9}{l}{\hspace{0.3cm}The ``effective" number of baselines is including only the telescopes where the burst emission falls significantly within the observing band. }\\
		\end{tabular}}
        
\end{table*}

\begin{table*}
\caption{\label{tab:burst_properties}Burst properties from Effelsberg single-dish data.}
\resizebox{\textwidth}{!}{\begin{tabular}{lcccccccc}
\hline
\hline
		{Burst}& {Time of Arrival$\mathrm{^{a}}$ } & {Fluence$\mathrm{^{b}}$} & {Peak S/N} & {Peak Flux Density$\mathrm{^{b}}$} & {Spectral Luminosity$\mathrm{^{c}}$} & {Width$\mathrm{^{d}}$ } & {Frequency Extent$\mathrm{^{d}}$ } & {Scintillation bandwidth$\mathrm{^{e}}$ } \\
		{} & {[MJD]} & {[Jy ms]} & &{[Jy]} & {[10$^{31}$\,erg\,s$^{-1}$\,Hz$^{-1}$]} & {[ms]} & {[MHz]} & {[MHz]}\\
		\hline
		Epoch 1\\
		B1 & 59314.63581536 & 0.94\,$\pm$\,0.19 & 7.1 & 0.31\,$\pm$\,0.06 & 2.38\,$\pm$\,0.48 & 4.0\,$\pm$\,0.3 & 141.1\,$\pm$\,0.5 & 1.60\,$\pm$\,0.63 \\ 
		B2 & 59314.67002510 & 3.68\,$\pm$\,0.74 & 11.4 & 0.73\,$\pm$\,0.15 & 6.78\,$\pm$1.36 & 5.6\,$\pm$\,0.1 & 120.2\,$\pm$\,0.1 & 1.01\,$\pm$\,0.34 \\ 
		B3 & 59314.69180857 & 0.61\,$\pm$\,0.12 & 4.9 & 0.24\,$\pm$\,0.05 & 1.34\,$\pm$\,0.27 & 4.7\,$\pm$\,0.1 & 107.1\,$\pm$\,0.7  & 1.86\,$\pm$\,1.42 \\ 
		B4 & 59314.71490874 & 1.49\,$\pm$\,0.30 & 9.0 & 0.43\,$\pm$\,0.09 & 2.56\,$\pm$\,0.51 & 6.0\,$\pm$\,0.1 & 98.4\,$\pm$\,0.2 & 1.27\,$\pm$\,0.59 \\ 
		B5 & 59314.75153586 & 1.53\,$\pm$\,0.31 & 9.2 & 0.46\,$\pm$\,0.09 & 0.33\,$\pm$\,0.67 & 4.7\,$\pm$\,0.2 & 97.2\,$\pm$\,0.3 & 1.14\,$\pm$\,0.59  \\ 
		B6 & 59314.79580654 & 6.50\,$\pm$\,1.30 & 13.4 & 0.91\,$\pm$0.16 & 8.20\,$\pm$\,1.64 & 8.1\,$\pm$\,0.1 & 131.4\,$\pm$\,0.1 & 0.95\,$\pm$\,0.29 \\ 
		B7 & 59314.80440853 & 0.94\,$\pm$\,0.19 & 7.7 & 0.52\,$\pm$\,0.10 & 4.11\,$\pm$\,0.82 & 2.4\,$\pm$\,0.1 & 107.3\,$\pm$\,0.2 &  1.55\,$\pm$\,0.54  \\ 
		B8 & 59314.82591958 & 4.48\,$\pm$\,0.90 & 13.2 & 0.92\,$\pm$\,0.18 & 7.29\,$\pm$\,1.26 & 6.3\,$\pm$\,0.1 & 89.6\,$\pm$\,0.1 & 0.89\,$\pm$\,0.37 \\ 
		B9 & 59314.83037838 & 1.40\,$\pm$\,0.28 & 6.3 & 0.27\,$\pm$\,0.05 & 2.40\,$\pm$\,0.48 & 5.9\,$\pm$\,0.1 & 81.2\,$\pm$\,0.1 & 1.00\,$\pm$\,0.49  \\ 
	    B10 & 59314.84159434 & 1.31\,$\pm$\,0.26 & 5.6 & 0.34\,$\pm$\,0.07 & 2.51\,$\pm$\,0.50 & 5.4\,$\pm$\,0.3 & 86.4\,$\pm$\,0.5  & 1.68\,$\pm$\,0.91  \\ 
	    B11 & 59314.85888953 & 5.92\,$\pm$\,1.18 & 14.6 & 0.89\,$\pm$\,0.18 & 11.4\,$\pm$\,2.3 & 5.3\,$\pm$\,0.1 & 155.3\,$\pm$\,0.1 & 0.65\,$\pm$\,0.27  \\ 
	    B12 & 59314.86388348 & 0.63\,$\pm$\,0.13 & 4.2 & 0.13\,$\pm$\,0.03 & 0.88\,$\pm$\,0.18 & 7.4\,$\pm$\,1.1 & 160.6\,$\pm$\,0.7 & - \\
		B13 & 59314.87198956 & 3.39\,$\pm$\,0.68 & 11.2 & 0.75\,$\pm$\,0.15 &  6.98\,$\pm$\,1.40 & 5.0\,$\pm$\,0.1 & 111.4\,$\pm$\,0.1 & 1.04\,$\pm$\,0.34 \\ 
		\hline
		Epoch 2\\
		B14$\mathrm{^{f}}$ & 59323.65617164 & 6.56\,$\pm$\,1.31 & 24.2 & 1.51\,$\pm$\,0.15 & 12.5\,$\pm$\,2.5 & 2.3\,$\pm$\,0.1 & 129.8\,$\pm$\,0.2 & 0.57\,$\pm$\,0.27  \\ 
		 & & & & & & 3.6\,$\pm$\,0.1 & 86.9\,$\pm$\,0.1 & \\
		B15 & 59323.66919992 & 1.98\,$\pm$\,0.40 & 20.2 & 1.23\,$\pm$\,0.25 & 12.6\,$\pm$\,2.5 & 1.6\,$\pm$\,0.1 & 136.7\,$\pm$\,0.1 & 1.23\,$\pm$\,0.34 \\ 
		B16 & 59323.71603797 & 1.95\,$\pm$\,0.39 & 8.2 & 0.5\,$\pm$\,0.1 & 4.15\,$\pm$\,0.83 & 4.8\,$\pm$\,0.1 & 125.5\,$\pm$\,0.1 & 1.49\,$\pm$\,0.53  \\ 
		B17 & 59323.74152501 & 1.78\,$\pm$\,0.36 & 7.2 & 0.31\,$\pm$\,0.06 & 2.94\,$\pm$\,0.59 & 6.1\,$\pm$\,0.1 & 143.6\,$\pm$\,0.2 & 0.55\,$\pm$\,0.41 \\ 
		B18 & 59323.79497897 & 0.91\,$\pm$\,0.18 & 5.5 & 0.24\,$\pm$\,0.05 & 2.43\,$\pm$\,0.49 & 3.9\,$\pm$\,0.1 & 170.8\,$\pm$\,0.6 & 1.10\,$\pm$\,1.07  \\ 
		
		\hline
		\multicolumn{9}{l}{$\mathrm{^{a}}$ Corrected to the Solar System Barycenter to infinite frequency assuming a dispersion measure of 412\,pc\,cm$^{-3}$,}\\
		\multicolumn{9}{l}{ \hspace{0.3cm} reference frequency 1502\,MHz and dispersion constant of 1/(2.41$\times 10^{-4}$)\,MHz$^2$\,pc$^{-1}$\,cm$^{3}$\,s.} \\
		\multicolumn{9}{l}{ \hspace{0.3cm} The times quoted are dynamical times (TDB).}\\
		\multicolumn{9}{l}{$\mathrm{^{b}}$ We estimate a conservative 20\% error on these measurements, arising due to the uncertainty in the system equivalent flux density (SEFD) of Effelsberg.} \\
		\multicolumn{9}{l}{$\mathrm{^{c}}$ Taking the luminosity distance of \frb\ as 453\,Mpc \citep{day_2021_atel, kilpatrick_2021_atel,hilmarsson_2021_mnras} }\\
		\multicolumn{9}{l}{$\mathrm{^{d}}$ Full-width at half-maximum of the Gaussian fit to the autocorrelation function of the dynamic spectrum.} \\
		\multicolumn{9}{l}{$\mathrm{^{e}}$ The uncertainty on the scintillation bandwidth is the quadrature sum of the fit errors and $1/\sqrt{N_{\sf scint}}$, where $N_{\sf scint}$ is the approximate number of scintles.} \\
		\multicolumn{9}{l}{$\mathrm{^{f}}$ B14 has two visible components. } \\
		\end{tabular}}
        
\end{table*}

\section*{Acknowledgements}
We would like to thank the directors and staff at the various participating stations for allowing us to use their facilities and running the observations. 
The European VLBI Network is a joint facility of independent European, African, Asian, and North American radio astronomy institutes. Scientific results from data presented in this publication are derived from the following EVN project code: EK048.
This work was also based on simultaneous EVN and PSRIX data recording observations with the 100-m telescope of the MPIfR (Max-Planck-Institut f\"{u}r Radioastronomie) at Effelsberg, and simultaneous EVN and DFB recording with the Sardinia 64-m telescope. We thank the local staff for this arrangement.
Research by the AstroFlash group at University of Amsterdam, ASTRON and JIVE is supported in part by an NWO Vici grant (PI Hessels; VI.C.192.045).
B.M. acknowledges support from the Spanish Ministerio de Econom\'ia y Competitividad (MINECO) under grant AYA2016-76012-C3-1-P and from the Spanish Ministerio de Ciencia e Innovaci\'on under grants PID2019-105510GB-C31 and CEX2019-000918-M of ICCUB (Unidad de Excelencia ``Mar\'ia de Maeztu'' 2020-2023).
Based on observations with the 100-m telescope of the MPIfR (Max-Planck-Institut für Radioastronomie) at Effelsberg.
JPY is supported by the National Program on Key Research and Development Project (2017YFA0402602).
NW acknowledges support from the National Natural Science Foundation of China (Grant 12041304 and 11873080).
The Sardinia Radio Telescope is funded by the Department of Universities and Research (MIUR), the Italian Space Agency (ASI), and the Autonomous Region of Sardinia (RAS), and is operated as a National Facility by the National Institute for Astrophysics (INAF).
This work is based in part on observations carried out using the 32-m Badary, Svetloe, and Zelenchukskaya radio telescopes operated by the Scientific Equipment Sharing Center of the Quasar VLBI Network (Russia).
This work is based in part on observations carried out using the 32-m radio telescope operated by the Institute of Astronomy of the Nicolaus Copernicus University in Toru\'n (Poland) and supported by a Polish Ministry of Science and Higher Education SpUBgrant.
%

\bibliographystyle{aasjournal}

\end{document}